\documentclass[12pt,journal,draftclsnofoot,onecolumn]{IEEEtran} 
\usepackage{amssymb}
\usepackage{amsmath}
\usepackage{amsmath,bm}
\usepackage{amsthm}
\usepackage{graphicx}
\usepackage{subfigure}
\usepackage{cite}
\usepackage{enumerate}

\begin{document}

\newtheorem{definition}{\bf ~~Definition}
\newtheorem{observation}{\bf ~~Observation}
\newtheorem{lemma}{\bf ~~Lamma}
\newtheorem{proposition}{\bf ~~Proposition}

\title{Radio Resource Allocation for Device-to-Device Underlay Communication Using Hypergraph Theory}
\author{
\IEEEauthorblockN{
{Hongliang Zhang}, \IEEEmembership{Student Member, IEEE},
{Lingyang Song}, \IEEEmembership{Senior Member, IEEE},
{and Zhu Han}, \IEEEmembership{Fellow, IEEE}}\\

\vspace{-0.5cm}

\thanks{H. Zhang and L. Song are with School of Electronics Engineering and Computer Science,
Peking University, Beijing, China (email: \{hongliang.zhang,
lingyang.song\}@pku.edu.cn).}

\thanks{Z. Han is with the Electrical and Computer Engineering
Department as well as Computer Science Department, University of Houston, Houston, USA (email:
zhan2@uh.edu).}
}

\maketitle

\vspace{-0.5cm}

\begin{abstract}

Device-to-Device (D2D) communication has been recognized as a promising technique to offload the traffic for the evolved Node B (eNB). However, the D2D transmission as an underlay causes severe interference to both the cellular and other D2D links, which imposes a great technical challenge to radio resource allocation. Conventional graph based resource allocation methods typically consider the interference between two user equipments (UEs), but they cannot model the interference from multiple UEs to completely characterize the interference. In this paper, we study channel allocation using hypergraph theory to coordinate the interference between D2D pairs and cellular UEs, where an arbitrary number of D2D pairs are allowed to share the uplink channels with the cellular UEs. Hypergraph coloring is used to model the cumulative interference from multiple D2D pairs, and thus, eliminate the mutual interference. Simulation results show that the system capacity is significantly improved using the proposed hypergraph method in comparison to the conventional graph based one.

\end{abstract}


\newpage
\section{Introduction}%

With the increasing demand for local traffic, device-to-device (D2D) communications under the control of evolved Node B (eNB) have recently received a great deal of attention~\cite{DRCCH-2009,JYDRWHTK-2009,GEGSNGZ-2012}. Reusing the same spectrum as for the cellular communications, user equipments (UEs) in a cellular network in proximity can set up direct transmissions, which potentially increases the overall spectral efficiency \cite{CLZ-2014}. In the Third Generation Partnership Project (3GPP), UEs are provided with a resource pool (time and frequency) in which they attempt to receive scheduling assignments, and eNB controls whether UEs may apply scheduled mode or autonomous mode D2D transmission \cite{3GPP-2014}. However, D2D communications generate interference to the cellular network if the radio resources are not properly allocated \cite{DLYGS-2013,PED-2013,MJPH-2011}. In addition, multiple D2D pairs in the same channel also create mutual interference \cite{LE-2008}. Thus, interference management becomes one critical issue for D2D communications underlaying cellular networks.

In the literature, much attention has been paid to manage the interference in D2D networks. The studies in \cite{HJBM-2011} propose a radio resource allocation algorithm using fractional frequency reuse to alleviate the interference between D2D pairs and cellular UEs. The work in \cite{FCLZB-2012,CLZQXB-2012} tackles the economy perspectives. In \cite{FCLZB-2012}, the authors formulate the allocation problem as a reverse iterative combinatorial auction game, and propose a joint radio resource and power allocation method to increase energy efficiency. In \cite{CLZQXB-2012}, a sequential second price auction mechanism is designed to allocate the spectrum resources for D2D communications with multiple user pairs.

As shown in the literature, though D2D communication may generate additional interference to cellular systems, it improves the system throughput with proper interference management~\cite{CKCO-2011}. Therefore, the allocation of radio resources for D2D underlay communications needs further studies for efficient solutions with low complexity. Graph theory is a useful tool to solve this kind of resource allocation problems in wireless communications \cite{HMKS-1999,AD-2013}. With graph theory, cellular UEs and D2D pairs are modeled as vertices in a graph, and the interference links between the UEs are constructed as edges \cite{HTLZ-2013,RXLB-2013}. In \cite{RXLB-2013}, the weight of the edges is used to represent the interference between two vertices, and the channel allocation is to iteratively gather vertices from the corresponding channel, taking both the interference value and the cluster value into account. In \cite{HTLZ-2013}, the system model is constructed as a weighted bipartite graph, and the channel allocation problem is formulated as a matching problem to maximize the capacity.

However, it is worth mentioning that the conception of edge in graph theory might not be sufficient in modeling the interference relation due to the cumulative effect of the interference. Specifically, the interference from several vertices may constitute a strong interferer, even though the interference from each individual vertex is weak \cite{QGR-2008,SK-1998}. When the cumulative interference from neighboring D2D pairs or cellular UEs exceeds a threshold, it may reduce the the communication quality of all the users. Hence, it is necessary to take into account the cumulative impact of multiple interference sources to the cellular UEs and D2D pairs as victims.

To this end, in this paper, we use the hypergraph to solve the interference management problem for D2D communication underlaying cellular networks. A hypergraph is a generalization of an undirected graph, in which the hyperedges are any subsets of the given set of vertices, instead of exactly two vertices defined in the traditional graph \cite{A-1974}. In wireless networks, the hypergraph achieves better approximation accuracy than the traditional graph as it effectively captures the cumulative interference. As such, the system capacity can be further improved by the hypergraph based method, compared to the traditional graph approach \cite{QR-2012}.

The main contributions of this paper are summarized as follows. We first formulate a resource allocation problem for multiple D2D pairs sharing channel resources with one cellular UE to maximize the cell capacity. Subsequently, we study the resource allocation problem using hypergraph theory. A hypergraph coloring method with low complexity is proposed to address the channel allocation for both D2D pairs and cellular UEs, which effectively increases the cell capacity. Simulation results show that the proposed hypergraph based method can achieve a performance very close to the optimal result, and performs much better than the traditional graph based method.

The paper is organized as follows. In Section \ref{system}, the D2D communications underlying cellular communication scenario is described, and the corresponding resource allocation problem is formulated. In Section \ref{graph}, we review a graph based channel allocation method. In Section \ref{hyper_graph}, a hypergraph based channel allocation method is proposed. In Section \ref{analysis}, the hypergraph based channel allocation method is analyzed and its complexity is compared to the graph based method. In Section \ref{simulation}, simulation results are provided. Finally in Section \ref{conclusion}, we draw the conclusions.

\section{System Model and Problem Formulation}%
\label{system}

\subsection{System Model}

\begin{figure}[!t]
\centering
\includegraphics[width=4.0in]{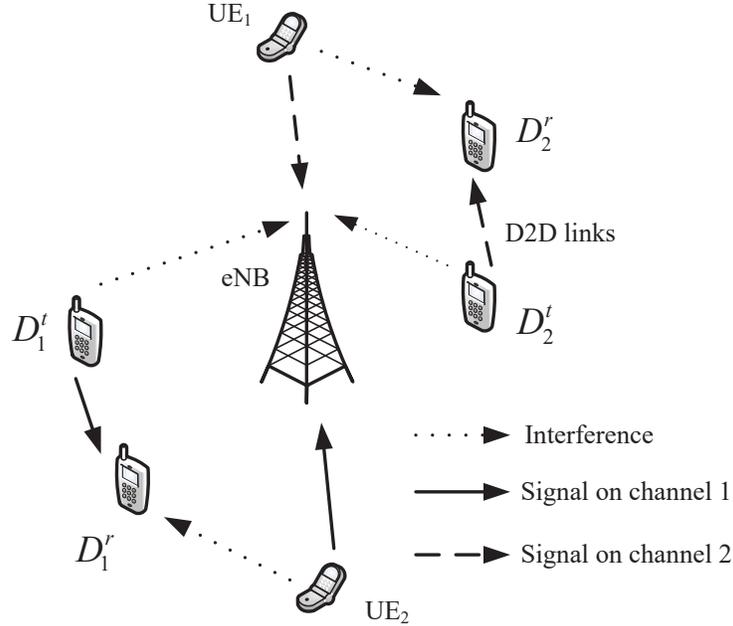}
\caption{System model for D2D communications underlaying cellular network when sharing uplink resource.} \label{system_model}
\end{figure}

As shown in Fig.~\ref{system_model}, we consider an uplink transmission scenario in a cellular network that consists of $N$ cellular UEs and $M$ D2D pairs. We denote a cellular UE by ${U_n}$, $1 \leq n \leq N$, and a D2D pair by $D_m$, $1 \leq m \leq M$. Here, we use ${D_m^t}$ to represent the transmitter of D2D pair $D_m$, and ${D_m^r}$ to represent the receiver of D2D pair $D_m$. Orthogonal Frequency Division Multiple Access (OFDMA) is employed to support multiple access for both the cellular and D2D communications, where a set of $K$ channels are available for resource allocation. In this system, the eNB coordinates the resource allocation between cellular UEs and D2D pairs. We assume that D2D pairs transmit with the power denoted by $P^d$, and cellular UEs use the transmission power $P^c$.

The channel is modeled as a Rayleigh fading channel, and the channel gains can be calculated by
\begin{equation}\label{Channel_gains}
\left\{ \begin{array}{ll}
{g_n^{c}} = L_n^{c} h_n^{c}, &\mbox{cellular link from}~{U_n}~\mbox{to eNB};\hfill\\
{g_m^{t,r}} = L_m^{t,r} h_m^{t,r}, &\mbox{D2D link from}~{D_m^t}~\mbox{to}~{D_m^r};\hfill\\
{g_m^{t}} = L_m^{t} h_m^{t}, &\mbox{link from}~{D_m^t}~\mbox{to eNB};\hfill\\
{g_{n,m}^{c,r}} = L_{n,m}^{c,r} h_{n,m}^{c,r}, &\mbox{link from}~{U_n}~\mbox{to}~{D_m^r};\hfill\\
{g_{i,m}^{t,r}} = L_{i,m}^{t,r} h_{i,m}^{t,r}, &\mbox{link from}~D_i^t~\mbox{to}~D_m^r,\hfill\\
\end{array} \right.
\end{equation}
where $L_n^{c}$, $L_m^{t,r}$, $L_m^{t}$, $L_{n,m}^{c,r}$, and $L_{i,m}^{t,r}$ denote the corresponding distance-dependent path loss, and $h_n^{c}$, $h_m^{t,r}$, $h_m^{t}$, $h_{n,m}^{c,r}$, and $h_{i,m}^{t,r}$ denote the fading channel, respectively, $1 \leq n \leq N$, $1 \leq m \leq M$, $1 \leq i \leq M$, and $i \ne m$. The thermal noise satisfies independent Gaussian distribution with zero mean and variance $ \sigma ^2$.

The instantaneous Signal to Interference-plus-Noise Ratio (SINR) of the received signal at the eNB from cellular UE $U_n$ in channel $k$ can be written as
\begin{equation}\label{Cellular_SINR}
\gamma _n^c = \frac{{{P^c}g_n^{c}}}{{{\sigma ^2} + \sum\limits_{m \in {\mathcal{C}_k}} {{P^d}g_{m}^{t}}}},
\end{equation}
and the instantaneous SINR at the D2D receiver $D_m^r$ in channel $k$ is given by
\begin{equation}\label{D2D_SINR}
\gamma _m^d = \frac{{{P^d}g_{m}^{t,r}}}{{{\sigma ^2} + \sum\limits_{{n} \in {\mathcal{C}_k}}{P^c}g_{n,m}^{c,r} + \sum\limits_{i \ne m, i \in {\mathcal{C}_k}} {{P^d}g_{i,m}^{t,r}} }},
\end{equation}
where $ \mathcal{C}_k$ represents the set of cellular UEs and D2D pairs to which channel $k$ is allocated.

\subsection{Problem Formulation}

We assume that a channel can be allocated to at most one cellular UE, and a maximum of one channel can be utilized by a D2D pair or a cellular UE. For convenience, we denote the channel allocation matrix by
\begin{equation}\label{Allocation_matrix}
S_{(N+M) \times K} = \left(\begin{array}{c} A_{N \times K}\\ B_{M \times K}\end{array}\right),
\end{equation}
where $A_{N \times K} = [\alpha_{n,k}]$ represents the channel allocation matrix for the cellular UEs, and \\ $B_{M \times K} = [\beta_{m,k}]$ stands for the channel allocation matrix for the D2D pairs, $1 \leq n \leq N$, $1 \leq m \leq M$, $1 \leq k \leq K$. The value of $\alpha_{n,k}$ and $\beta_{m,k}$ are defined as
\begin{equation}\label{D_alpha}
\alpha_{n,k}=
\left\{ \begin{gathered}
1,~~\mbox{when channel}~k~\mbox{is allocated to} ~U_n,\hfill\\
0,~~\mbox{otherwise},\hfill\\
\end{gathered} \right.
\end{equation}
and
\begin{equation}\label{D_beta}
\beta_{m,k}=
\left\{ \begin{gathered}
1,~~\mbox{when channel}~k~\mbox{is allocated to} ~D_m,\hfill\\
0,~~\mbox{otherwise}.\hfill\\
\end{gathered} \right.
\end{equation}

Our objective is to maximize the cell capacity by optimizing the channel allocation variables $\{\alpha_{n,k}; \beta_{m,k}\}$ for the cellular UEs and D2D pairs, which can be formulated as
\begin{equation}\label{Objective}
\max
\hspace{-0.15cm}\sum\limits_{k = 1}^K {\left[\sum\limits_{n = 1}^N {{{\log }_2}(1\hspace{-0.12cm}+\hspace{-0.12cm} \gamma _n^c)} \alpha_{n,k}
\hspace{-0.15cm}+
\hspace{-0.15cm} \sum\limits_{m = 1}^M {{{\log }_2}(1+ } \gamma _m^d) \beta_{m,k} \right]}
\end{equation}
\begin{equation}\label{Constraint}
st.~~
\left\{ \begin{gathered}
\sum\limits_{n = 1}^N {{\alpha _{n,k}}}  \leq 1,\hfill\\
\sum\limits_{k = 1}^K {{\alpha _{n,k}}}  \leq 1,~~\sum\limits_{k = 1}^K {{\beta _{m,k}}} \leq 1,\hfill\\
\end{gathered} \right.
\end{equation}
where $\gamma _n^c$ and $\gamma _m^d$ are given in (\ref{Cellular_SINR}) and (\ref{D2D_SINR}), respectively. Constraints in (\ref{Constraint}) imply that each channel can be allocated to at most one cellular UE, and a maximum of one channel can be utilized by each D2D pair or each cellular UE.

 Note that the aforementioned resource allocation problem in (\ref{Objective}) is a NP-hard combinatorial optimization problem with nonlinear constraints \cite{D-1976}, graph coloring is an approximate and efficient method for such a resource allocation problem \cite{RZJC-2008}. Thus, we formulate the channel resources as $K$ different colors, the cellular UEs as $N$ (cellular) vertices, and the D2D pairs as $M$ (D2D) vertices in the plane. Consequently, the channel allocation problem is transformed into a coloring problem of the vertices with fixed colors \cite{TB-1995}. In the following two sections, we demonstrate the graph and the hypergraph based methods, respectively.

\section{Traditional Graph Based Channel Allocation}%
\label{graph}

\begin{table}[!t]
\renewcommand{\arraystretch}{1.0}
\caption{Algorithm \uppercase\expandafter{\romannumeral1}: Graph Based Resource Allocation Method} \label{graph_algorithm} \centering
\begin{tabular}{p{115mm}}
\hline

\textbf{\emph{Stage \uppercase\expandafter{\romannumeral1}: Graph Construction}}  \\
\quad $\ast$ Cellular UEs $U_n$ and $U_j$ form an edge, $\forall U_n$, $U_j$, where $n \neq j$.\\
\quad $\ast$ A cellular UE $U_n$ and a D2D pair $D_m$ form an edge if they satisfy (\ref{graph_interference_DC}) or (\ref{graph_interference_CD}).\\
\quad $\ast$ D2D pairs $D_i$ and $D_m$ form an edge if they satisfy (\ref{graph_interference_DD}).\\
\textbf{\emph{Stage \uppercase\expandafter{\romannumeral2}: Graph Coloring Algorithm}}  \\
\quad $\ast$ $i=1$. Find a vertex of the maximum degree and label it $x_i$.\\
\quad $\ast$ \textbf{repeat}
\begin{enumerate}
    \item Set $i = i + 1$. Select from the unexamined subgraph a vertex $x$ which has the maximum degree, and label it $x_i$.
    \item Break the edges which connect to vertex $x_i$;
\end{enumerate}
\quad $\ast$ \textbf{until} All the vertices in the graph are examined. \\
\quad $\ast$ Starting from $i = 1$, select a color randomly from the available color set to color $x_i$. If the available color set is empty, leave the vertex $x_i$ uncolored.\\
\hline
\end{tabular}
\end{table}

Before introducing the hypergraph based channel allocation method, we describe the conventional graph based method.

\begin{definition}\label{graphdefinition}
A graph $G$ is defined to be a pair $(X,E)$, where $X = \{x_1,x_2,\ldots,x_n\}$ is a set of elements called vertices, and $E = \{e_1,e_2,\ldots,e_m \}$ is a set of 2-element subsets of $X$ called edges.
\end{definition}

In a graph, vertices represent the cellular UEs and the D2D pairs, and edges indicate that the interference between connected vertices does not allow them to use the same channel simultaneously \cite{DENL-2012}. The graph based method contains the graph construction and the channel allocation algorithm as follows.

\subsubsection{Graph Construction}

 We transform the interference information into a graph. A cellular UE $U_n$ and a D2D pair $D_m$ are connected by
 an edge which satisfies that the wanted signal ratio to the interference is below a threshold:
\begin{equation}\label{graph_interference_DC}
\frac{P^c{g_n^{c}}}{P^d{g_m^{t}}} < {\delta _c};~\mbox{at the eNB receiver},
\end{equation}
or
\begin{equation}\label{graph_interference_CD}
\frac{P^d{g_m^{t,r}}}{P^c{g_{n,m}^{c,r}}} < {\delta _d};~\mbox{at the D2D receiver}~D_m,
\end{equation}
where $\delta _c$ and $\delta _d$ are the thresholds selected to determine the severity of the interference at the eNB and the receiver of a D2D pair, respectively. Two D2D pairs $D_i$ and $D_m$ are connected by an edge if
\begin{equation}\label{graph_interference_DD}
\frac{{g_m^{t,r}}}{{g_{i,m}^{t,r}}} < {\delta _d};~\mbox{at the D2D receiver}~D_m,
\end{equation}
which indicates that if the interference from another D2D pair is strong, these two D2D pairs cannot share the same channel. Besides, two cellular UEs $U_i$ and $U_j$ always form an edge for the assumption that two cellular UEs cannot share the same channel. In this way, an interference graph is constructed.

\subsubsection{Channel Allocation Algorithm}

After the graph construction, we use the greedy coloring algorithm in \cite{RZJC-2009} to color the constructed graph. We define the available color set by all the colors except the colors used in the connected vertices. The algorithm successively colors the vertices in a color randomly chosen in the corresponding available color set, in descending order of degree. If the available color set becomes empty, the vertex remains uncolored. In this way, the cellular UEs and the D2D pairs are classified into clusters with different colors, where the colors represent the channels. Finally, the channels are allocated to the D2D pairs and cellular UEs with mutual interference below the given threshold. These detailed algorithms are shown in Table \ref{graph_algorithm}.

\section{Hypergraph Based Channel Allocation}%
\label{hyper_graph}

In the traditional graph based method of Section \ref{graph}, the edge connecting two vertices is not sufficient to model the interference in a wireless network, because some weak interferers together may constitute a strong cumulative interferer to affect the link quality. In this section, the hypergraph method, in which a hyperedge contains several vertices, is used for interference modelling.

\subsection{Hypergraph Preliminaries}

Before proposing the hypergraph based channel allocation method, we first introduce some preliminaries of hypergraph theory \cite{C-1973}. Hypergraph is a generalized graph, in which edges consist of any subset of the given set of vertices instead of exactly two vertices defined in the traditional graph.

\begin{definition}\label{hypergraphdefinition}
Let $X = \{x_1,x_2,\ldots,x_n\}$ be a finite set, a hypergraph $H$ on $X$ is a family $E = (e_1,e_2,\ldots,e_m)$ of subsets of $X$ such that
\begin{equation}\label{Hypergraph_definition}
\begin{split}&
e_i \neq \emptyset~(i = 1,2,\ldots,m),\\
&\bigcup\limits_{i = 1}^m {e_i }  = X.
\end{split}
\end{equation}
The elements $x_1,x_2,\ldots,x_n$ of $X$ are vertices of hypergraph $H$, and the sets $e_1,e_2,\ldots,e_m$ are the hyperedges of hypergraph $H$.
\end{definition}

\begin{figure}[!t]
\centering
\includegraphics[width=4.0in]{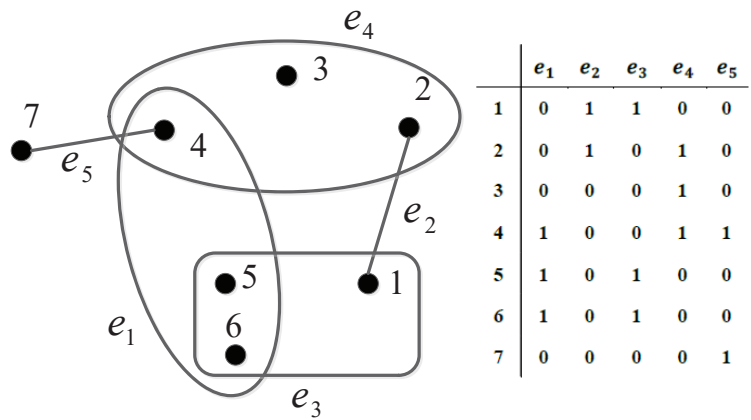}
\caption{An example of hypergraph $H$ and its incidence matrix.} \label{hypergraph}
\end{figure}

The traditional graph can be specified from its incidence matrix or adjacency matrix \cite{C-1985}. The incidence matrix has one row for each vertex and one column for each edge. If vertex $x_i$ is incident to edge $e_j$, then $(i,j)$-entry in the matrix is 1, otherwise it is 0. The adjacency matrix has one row and one column for each vertex. If vertex $x_i$ is adjacent to vertex $x_j$, then $(i,j)$-entry in the matrix is 1, otherwise it is 0. However, different from the traditional graph, there does not exist one-to-one correspondence between a hypergraph and its adjacency matrix, and only the incidence matrix can determine a hypergraph. The edge set in which all the edges contain vertex $x$ is represented by $E(x)$. The degree of vertex $x$ can be then defined as the cardinality of $E(x)$, denoted by $|E(x)|$. The traditional graph is a hypergraph in which the degree of vertices is always 2. A simple example of a hypergraph is given in Fig.~\ref{hypergraph}, where the left figure is a hypergraph with five edges and the right table is its corresponding incidence matrix. For instance, as shown in Fig.~\ref{hypergraph}, the hyperedge $e_3$ contains $x_1$, $x_5$ and $x_6$, and in the incident matrix, elements $(1,3)$, $(5,3)$, and $(6,3)$ are 1.

\subsection{Hypergraph Construction}

In this subsection, we will present a hypergraph based channel allocation method to solve the resource allocation problem. The first step is to construct the hypergraph for the mutual interference between D2D pairs and cellular UEs, and the next one is to color the constructed hypergraph. By hypergraph coloring, different subsets of cellular UEs and D2D pairs are generated, where one subset corresponds to one channel. Finally, orthogonal channels are assigned to each subset, which means that the cellular UE and D2D pairs in the subset share the channel.

In the hypergraph construction, we define two kinds of interferers. The first kind is \emph{independent interferer}, and the second one is \emph{cumulative interferer}. We define that the independent interferers of a D2D receiver or the eNB receiver are the D2D pairs and cellular UEs which decrease the received SINR independently. The cumulative interferers decrease SINR notably when combined in the receiver. We construct the hypergraph by the following steps:

\subsubsection{Independent Interferer Recognition}

The first step is to select the independent interferers. Under the assumption that a maximum of one channel can be utilized by a cellular UE, one cellular UE can be regarded as an independent interferer of another, and thus, they form an edge. This step is to avoid the severe interference which originates from two UEs sharing the same channel. We give an example in Fig.~\ref{illustration} with three cellular UEs and three D2D pairs which are denoted by $U_1$, $U_2$, $U_3$, $D_1$, $D_2$ and $D_3$, respectively. According to the aforementioned construction, cellular $U_1$, $U_2$ and $U_3$ form edge 5, edge 6 and edge 7.

Next, we search the independent interferers for each UE, and construct the corresponding edges. Similar to the graph based method, for the cellular UEs, we follow the pairwise comparison as we have done in Section \ref{graph} to select the independent interferers. If cellular UE $U_n$ and D2D pair $D_m$ satisfy (\ref{graph_interference_DC}) or (\ref{graph_interference_CD}), they form an edge. Similarly, we also make the pairwise comparison for the D2D pairs to select independent interferers. If D2D pairs $D_i$ and $D_m$ satisfy (\ref{graph_interference_DD}), they form an edge as well. As shown in Fig.~\ref{illustration}, $U_1$ and $D_1$ form edge 1, and $U_3$ and $D_1$ form edge 2. In the next paragraph, we construct the hyperedges, accounting for the cumulative interference from different users.

\subsubsection{Cumulative Interferer Recognition}

\begin{figure}[!t]
\centering
\includegraphics[width=4.0in]{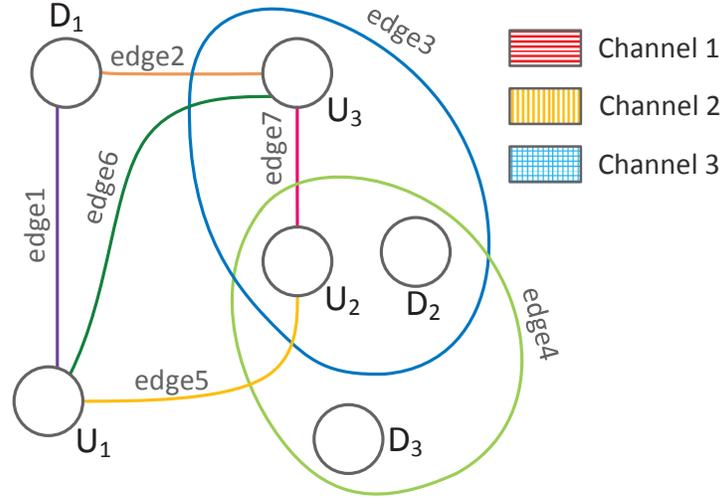}
\caption{An example of the hypergraph modeling.} \label{illustration}
\end{figure}

After all the independent interferers are determined, the next step is to find the cumulative interferers, and construct the hyperedge. The cumulative interference is gathered from more than one UEs, except the independent interferers. We select a number of UEs, and compare the cumulative interference with an interference threshold $\eta$ to verify whether they become interferers if cumulated. For instance, we select $Q$ UEs, including cellular and D2D interferers, and then compare the cumulative interference to the wanted signal to determine whether they together form a hyperedge. For a cellular UE $U_n$, if the wanted signal to the cumulative interference ratio is below a threshold $\eta_c$, the cumulative interferers and the cellular UE together form a hyperedge, i.e.,
\begin{equation}\label{hypergraph_interference_CD}
\frac{P^c{g_n^{c}}}{\sum\limits_{m = 1}^{{G}} {P^d g_m^{t}} } < {\eta _c};~\mbox{at the eNB receiver}.
\end{equation}
And for a D2D pair $D_m$, if the wanted signal to the cumulative interference ratio is below a threshold $\eta_d$, the cumulative interferers and the D2D pair together form a hyperedge, i.e.,
\begin{equation}\label{hypergraph_interference_DCD}
\frac{P^d{g_m^{t,r}}}{{\sum\limits_{j = 1}^{{F_m}} {P^c g_{j,m}^{c,r}} } +{\sum\limits_{i = 1}^{{Z_m}} {P^d g_{i,m}^{t,r}} }} < {\eta _d};~\mbox{at the D2D receiver}.
\end{equation}
Here, $F_m$ and $Z_m$ are the number of the cellular and D2D interferers in the hyperedge, respectively, i.e, $Z_m+F_m = Q$. As the example shown in Fig.~\ref{illustration}, $U_2$, $D_2$ and $D_3$ form edge 4, and $U_2$, $U_3$ and $D_2$ form edge 3.

It is worth mentioning that the value of $Q$ is optional. Here, we only consider constructing the hypergraph with $Q$ equal to 2, because it would be sufficient for the modeling. The hyperedge is to select $Q$ UEs which generate severe interference to the examined UE, and judge whether the interference meets the criteria. With a higher value of $Q$, the complexity of the construction will increase. However, from the simulation results in Section \ref{simulation}, the cell capacity will increase less than 1$\%$ when the value of $Q$ adds by 1. To achieve a compromise between cell capacity and computational complexity, we construct the hypergraph with $Q = 2$.

By definition, the union of hyperedges need to be the vertex set $X$. A special case may occur, where one vertex is neither an independent interferer of any UE, nor any of the cumulative interferers. In such a case, the union of hyperedges is not equal to the vertex set $X$. The vertex which is not in any hyperedge forms a hyperedge itself. In this way, the union of hyperedges is equal to vertex set $X$. After all these steps, hypergraph $H$ can be constructed.

\subsection{Hypergraph Coloring Algorithm}

\begin{table}[!t]
\renewcommand{\arraystretch}{1.0}
\caption{Algorithm \uppercase\expandafter{\romannumeral2}: Hypergraph Based Resource Sharing Method} \label{algorithm} \centering
\begin{tabular}{p{115mm}}
\hline
\textbf{\emph{Stage \uppercase\expandafter{\romannumeral1}: Hypergraph Construction}}  \\
\quad $\ast$ One cellular UE can be regarded as the independent interferer of another, and thus, two cellular UEs form an edge.\\
\quad $\ast$ \textbf{repeat}
\begin{enumerate}
    \item Compare the SINR with the threshold $\delta$ to select independent interferers. For a cellular UE $U_n$, if it satisfies (\ref{graph_interference_DC}), the D2D pair $D_m$ is an independent interferer. And for a D2D pair $D_m$, if it satisfies  (\ref{graph_interference_CD}) or (\ref{graph_interference_DD}), the cellular UE $U_n$ or D2D pair $D_i$ is an independent interferer as well.
    \item Form edges with the independent interferers.
\end{enumerate}
\quad $\ast$ \textbf{until} All UEs find their independent interferers.\\
\quad $\ast$ \textbf{repeat}
\begin{enumerate}
    \item Compare the SINR with the threshold $\eta$ to find cumulative interferers. For a cellular UE $U_n$, if it satisfies (\ref{hypergraph_interference_CD}), the D2D pairs are the cumulative interferers. And for a D2D pair $D_m$, if it satisfies (\ref{hypergraph_interference_DCD}), the cellular UEs and D2D pairs are the cumulative interferers.
    \item Form hyperedges with the cumulative interferers.
\end{enumerate}
\quad $\ast$ \textbf{until} All UEs find their cumulative interferers.\\
\quad $\ast$ The vertex which is not in any hyperedge or edge forms a hyperedge itself.\\
\textbf{\emph{Stage \uppercase\expandafter{\romannumeral2}: Hypergraph Coloring Algorithm}}  \\
\quad $\ast$ $i=n$, $H_n = H$. Find a vertex of the minimum monodegree in $H_n$ and label it $x_n$.\\
\quad $\ast$ \textbf{repeat}
\begin{enumerate}
    \item Set $i = i-1$, and strongly delete the vertex $x_{i+1}$ and form an induced sub-hypergraph $H_i = H_{i+1}-x_{i+1}$.
    \item Find a vertex of the minimum monodegree in $H_i$ and label it $x_i$.
\end{enumerate}
\quad $\ast$ \textbf{until} $i=0$. \\
\quad $\ast$ Starting from $i = 1$, color the vertex $x_i$ in a color randomly selected from the corresponding available color set, successively. When the available color set is empty, remain the vertex $x_i$ uncolored. \\
\hline
\end{tabular}
\end{table}

After hypergraph construction, hypergraph $H$ can be colored. A color in the hypergraph corresponds to a channel, and coloring vertices is equivalent to allocating a channel to the D2D pairs and cellular UEs. Similar to the graph coloring in Section \ref{graph}, the vertices contained in the same hyperedge cannot be colored by the same color. In this way, the cumulative interference can be alleviated.

Since coloring of the hypergraph is NP-hard, there is no computationally efficient algorithm to obtain the optimal solution \cite{V-2008}. Coloring algorithms have been proposed to color a hypergraph efficiently in \cite{V-2002}. The one mentioned in \cite{V-2008} is a greedy algorithm to color the hypergraph which is colorable. This implies that there exists a sufficient number of colors to color the hypergraph. However, in the OFDMA network, the condition may not be fulfilled, because the number of vertices may change as a function of cell load, while the number of channels is fixed. If the network is heavily loaded, it is not possible to color the whole hypergraph. In the light of these observations, we propose to modify the greedy method mentioned in \cite{V-2008} to meet the needs in an OFDMA cell. The necessary definitions are formulated below.

\begin{definition}\label{deletion}
In a hypergraph $H(X,E)$, \emph{strong deletion} of a vertex $x \in X$ from $H$ is to delete all the edges containing $x$ from $E$, and delete $x$ from $X$.
\end{definition}

\begin{definition}\label{subhyperedge}
A hypergraph $H^{'}(X^{'},E^{'})$ is called \emph{sub-hypergraph} of a hypergraph $H(X,E)$ if $X^{'} \subseteq X$, and $E^{'} \subseteq E$. And the sub-hypergraph $H^{'}(X^{'},E^{'})$ is called \emph{induced sub-hypergraph} when the hyperedges of $H(X,E)$ completely contained in $X^{'}$ form the hyperedge set $E^{'}$ .
\end{definition}

An induced sub-hypergraph $H^{'}$ is a special case of sub-hypergraphs, which can be obtained from $H$ by strong deletion of vertices $X-X^{'}$. If at least one hyperedge of $H$ being a subset of $X^{'}$ is empty, the sub-hypergraph is not induced.

\begin{definition}\label{monodegree}
The monodegree $m(x,H)$ \cite{V-2008} of vertex $x \in X$ in a hypergraph $H(X,E)$ is the maximum cardinality of a hyperedge subfamily $E_1(x) \subseteq E(x)$ such that two elements $e_i,e_j \in E_1(x)$, $e_i \cap e_j = \{x\}$.
\end{definition}

In other words, the monodegree of vertex $x$ is the maximum size of such a hyperedge set, where every two hyperedges share precisely one vertex $x$. Intuitively speaking, the hyperedge set looks like a star, where vertex $x$ is in the center of the star. If a graph has no loops, which implies that the two vertices in an edge are not the same, the monodegree is equal to the degree in the graph. We consider the value $M(H) = \mathop {\max }\limits_{Y \subseteq X} \mathop {\min }\limits_{x \in Y} m(x,H \backslash Y)$. It can be obtained by selecting a vertex of the minimum monodegree, and making the monodegree maximum over all the induced sub-hypergraphs. The value $M(H)$ is related to the minimum number of colors needed when the hypergraph is totally colored. This property will be further discussed in Section \ref{analysis}.

The modified method is presented in Table \ref{algorithm}. The difference between this modified method and the greedy method in \cite{V-2008} lies in the number of colors. The modified method uses a fixed number of colors instead of the lowest number of colors in \cite{V-2008}. According to Algorithm \uppercase\expandafter{\romannumeral2}, the D2D pairs and the cellular UEs have equivalent opportunity in resource allocation. When the D2D pairs have better channel conditions, the D2D pairs can be allocated to channels instead of the cellular UEs.

It is worth mentioning that hypergraph coloring is a method to obtain the sub-optimal solution in polynomial time. According to the description of Algorithm \uppercase\expandafter{\romannumeral2}, the vertex with maximum monodegree is colored first. This implies that the UE which generate largest interference are allocated to the channels first, then other UEs can utilize other channels to avoid the interference. In this way, more UEs can be allocated to channels, and hence the capacity increases. Hypergraph coloring is therefore a greedy method to obtain a sub-optimal solution.

\section{Theoretical Analysis}%
\label{analysis}

In this section, we first evaluate the performance of the hypergraph based method, and then address the computational complexity of both the graph and the hypergraph methods.

\subsection{Property Analysis}

For the comparison of these two methods, we provide the following propositions.

\begin{proposition}\label{saturated}
When the number of cellular UEs and the number of channels are fixed, the cell capacity will first increase and then become saturated as the number of D2D pairs increases.
\end{proposition}

\begin{IEEEproof}
For a large number of D2D pairs, we assume that the monodegree of D2D pair $x$ is the lowest. If the monodegree of D2D pair $x$ is higher than the number of colors $K$, the D2D pair $x$ cannot be colored, i.e., allocated to the channels. When the number of D2D pairs grows, the traditional graph and the hypergraph methods only select those D2D pairs, which generate less interference to replace the previous candidates. This is the reason why the capacity becomes saturated with the increasing number of D2D pairs.
\end{IEEEproof}

\begin{observation}\label{minimummonodegree}
The maximum value of the minimum monodegree generated by Algorithm \uppercase\expandafter{\romannumeral2} is equal to $M(H)$.
\end{observation}

\begin{IEEEproof}
According to \textbf{Definition \ref{monodegree}}, the maximum value of the minimum monodegree over all vertices $d$ generated by Algorithm \uppercase\expandafter{\romannumeral2} needs to satisfy that $d \leqslant M(H)$. On the other hand, Algorithm \uppercase\expandafter{\romannumeral2} strongly deletes the vertex of the minimum monodegree, and there must be an induced sub-hypergraph $H \backslash Y_0$ obtained by also strongly deleting those vertices in $Y_0$. For a vertex $y \in Y_0$,
\begin{equation}
m(y,H \backslash {Y_0}) = \mathop {\min }\limits_z m(z,H \backslash {Y_0}) = M(H).
\end{equation}
In the generic step $l \geq 1$ of Algorithm \uppercase\expandafter{\romannumeral2}, the first vertex is deleted from set $Y_0$ such that the minimum monodegree of the induced hypergraph $H \backslash Y_0$ is equal to $M(H)$. Thus, $H \backslash Y_0$ is an induced sub-hypergraph of $H_l$. The minimum monodegree $m(x_l,H_l)$ of $H_l$ is higher than that of $H \backslash Y_0$. Therefore,
\begin{equation}
M(H) = m(y,H \backslash Y_0) \leq m(x_l,H_l) \leq d.
\end{equation}
\end{IEEEproof}

\begin{proposition}\label{color}
The minimum number of colors to make all the vertices in hypergraph $H$ colored is defined by $X(H)$, and $X(H) = M(H) + 1$.
\end{proposition}

\begin{IEEEproof}
From Observation \ref{minimummonodegree}, the upperbound of the minimum monodegree obtained by Algorithm \uppercase\expandafter{\romannumeral2} is equal to $M(H)$. In the coloring process, vertex $x_i$ will be in at most $M(H)$ hyperedges. In the case where the vertices in these hyperedges are colored differently, the number of required colors is largest. Thus, in the coloring process, the number of colors used is not less than $M(H)$. In addition, these hyperedges have the unique common vertex $x_i$. Thus, the next new color is needed for this vertex $x_i$.
\end{IEEEproof}

\textbf{Proposition \ref{color}} indicates that if the number of the channels is larger than $M(H)$, all the cellular UEs and D2D pairs can be allocated to channels.

\begin{proposition}\label{topology}
We assume the vertex set $X$ of hypergraph $H$ is divided into cellular set $X_c$ and D2D set $X_d$. When the number of cellular UEs increases by 1, the cellular UEs and D2D pairs form a new hypergraph $H^{'}$. If $M(H) = \mathop {\max }\limits_{Y \subseteq X_c} \mathop {\min }\limits_{x \in Y} m(x,H \backslash Y)$, then $M(H^{'}) = M(H) + 1$; Otherwise, $M(H) \leq M(H^{'}) \leq M(H) + 1$.
\end{proposition}

\begin{IEEEproof}
 In hypergraph construction, if the number of vertices increases by 1, the monodegree of the other vertices will increase by at most 1. The reason is that once two vertices form an edge, one vertex will not be the cumulative interferer of the other, and they cannot form a hyperedge. In addition, any two cellular UEs are bound to form an edge, and thus, if the number of cellular UEs increases by 1, the monodegree of each cellular UE will increase by 1 as well.

Under the assumption $M(H) = \mathop {\max }\limits_{Y \subseteq X_c} \mathop {\min }\limits_{x \in Y} m(x,H \backslash Y)$, cellular UE $x$ is the vertex which has the maximum value of the minimum monodegree. According to the aforementioned analysis, if the monodegree of cellular UE $x$ increases by 1, then the monodegree of the other vertices will increase by at most 1. Thus, cellular UE $x$ is still the vertex which has the maximum value of the minimum monodegree, and $M(H^{'}) = M(H) + 1$. Otherwise, a D2D pair $x$ is the vertex which has the maximum value of the minimum monodegree. If the mutual interference between D2D pair $x$ and the new cellular UE cannot form an edge nor a hyperedge, $M(H^{'}) = M(H)$. Therefore, if the vertex is not a cellular UE, $M(H) \leq M(H^{'}) \leq M(H) + 1$.
\end{IEEEproof}

\subsection{Complexity Analysis}

According to Algorithm \uppercase\expandafter{\romannumeral1}, the graph based resource allocation method can be processed in a greedy manner. For the graph based method, the complexity of calculating the interference of the D2D pairs and cellular UEs is proportional to $O(MN + N^2)$. For graph coloring, it is necessary to go through all the vertices and break at most $(M+N)(M+N-1)$ edges. The computational complexity of the graph based channel allocation is quadratic given by
\begin{equation}\label{c_graph}
C_G \propto O((M+N)^2).
\end{equation}

According to Algorithm \uppercase\expandafter{\romannumeral2}, the hypergraph based resource allocation method is processed in a greedy manner as well. For the hypergraph based method, the complexity of finding the independent interferers is equal to the graph based method, i.e., proportional to $O(MN + N^2)$. The complexity of finding the cumulative interferers of the D2D pairs and cellular UEs is proportional to $O((M+N)^2)$. For hypergraph coloring, there exist at most $(M+N-1)$ two-verticed edges and $(M+N-1)(M+N-2)$ hyperedges, and the method requires going through all the vertices and breaking at most $((M+N)(M+N-1)(M+N-2))$ edges. The computational complexity of the hypergraph based channel allocation method is cubic given by
\begin{equation}\label{c_hypergraph}
C_H \propto O((M+N)^3).
\end{equation}

From this analysis, we can conclude that the hypergraph based channel allocation method takes cubic polynomial time, in comparison to the graph based channel allocation method, which takes quadratic polynomial time.

\section{Simulation Results}%
\label{simulation}

\begin{table}
\centering
\caption{Parameters for Simulation} \label{parameters}
\begin{tabular}{|p{4.5cm}|p{4.5cm}|}
 \hline Cellular layout & Isolated cell\\
 \hline Cell Radius & 500 m\\
 \hline Maximum D2D Pair Distance & 20 m \\
 \hline Cellular UE's Transmit Power $P^c$ & 23 dBm\\
 \hline D2D's Transmit Power $P^d$ & 13 dBm \\
 \hline Carrier Frequency & 2.3 GHz \\
 \hline Transmission Bandwidth & 20 MHz \\
 \hline Noise Figure & 5 dB \\
 \hline Threshold $\delta_c = \eta_c$ & 20 dB \\
 \hline Threshold $\delta_d = \eta_d$ & 20 dB \\
 \hline Path Loss Model & UMi in \cite{ITU-2008} \\
 \hline Small Scale Fading & Rayleigh fading coefficient with zero mean and unit variance \\
\hline
\end{tabular}
\end{table}

\begin{figure}[!t]
\centering
\includegraphics[width=5in]{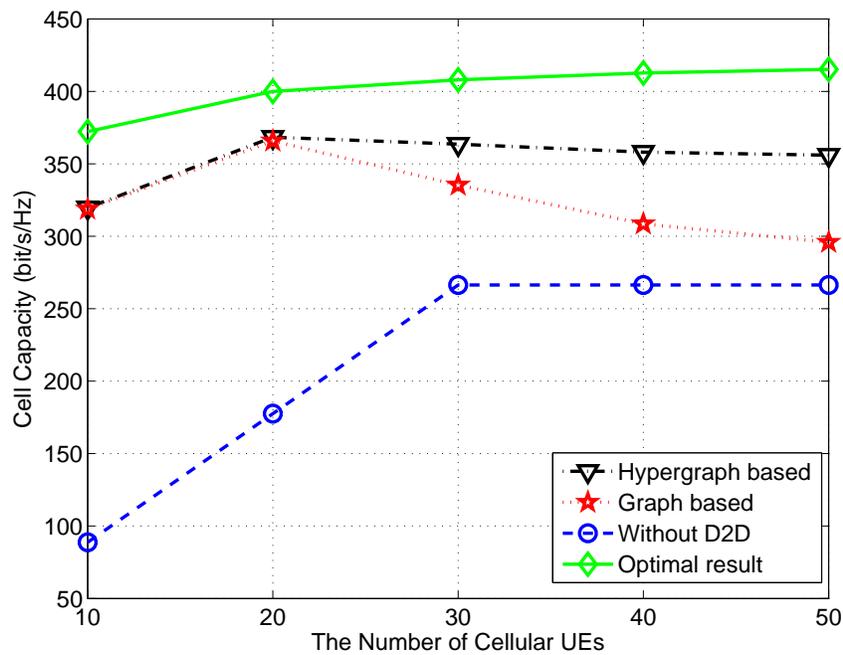}
\caption{The cell capacity with the number of cellular UEs $N$ for $K = 30$, and $M = 20$.} \label{c_cellular}
\end{figure}

\begin{figure}[!t]
\centering
\includegraphics[width=5in]{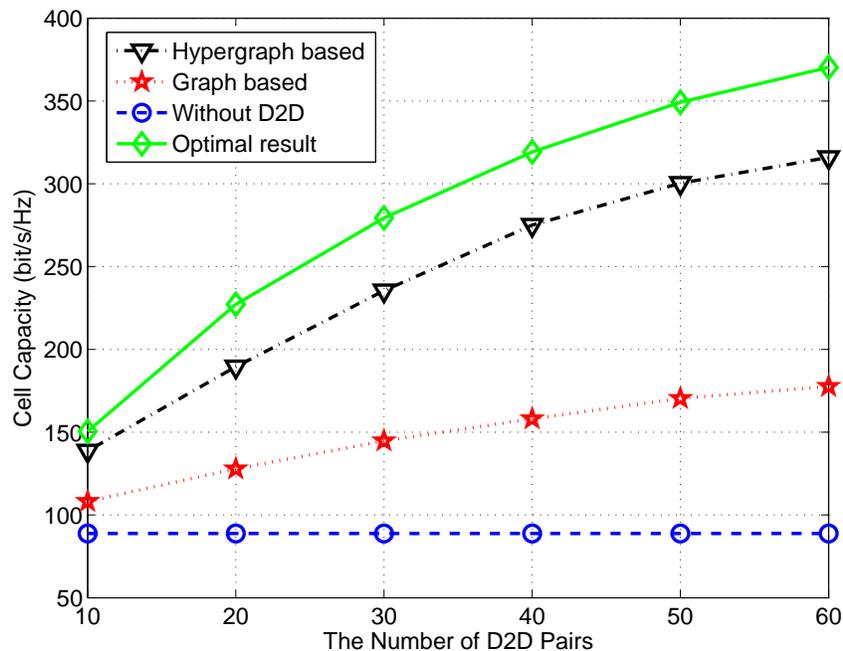}
\caption{The cell capacity with the number of D2D pairs $M$ for $K = 10$, and $N = 10$.} \label{c_D2D}
\end{figure}

\begin{figure}[!t]
\centering
\includegraphics[width=5in]{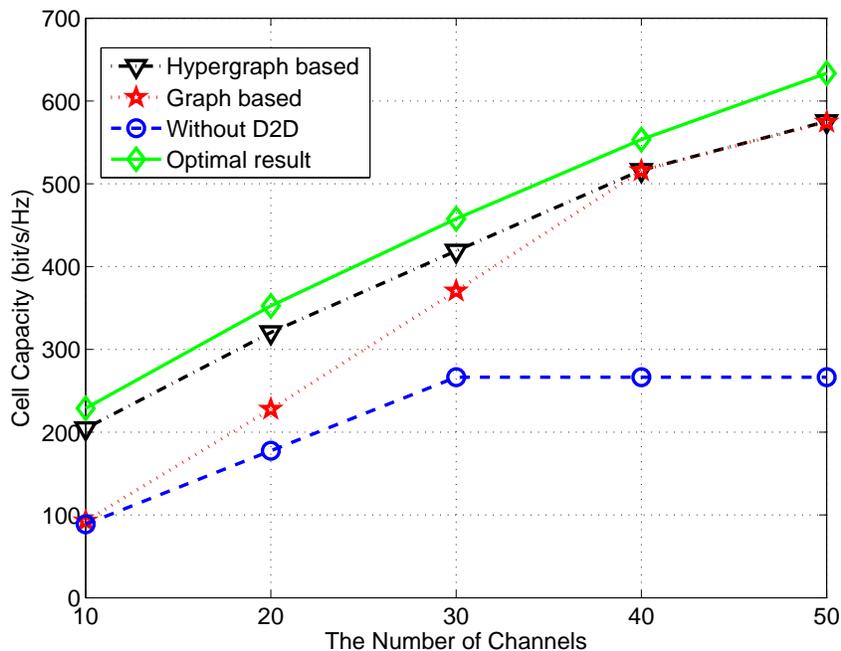}
\caption{The cell capacity with the number of channels $K$ for $M = 30$, and $N = 30$.} \label{c_channel}
\end{figure}

\begin{figure}[!t]
\centering
\includegraphics[width=5in]{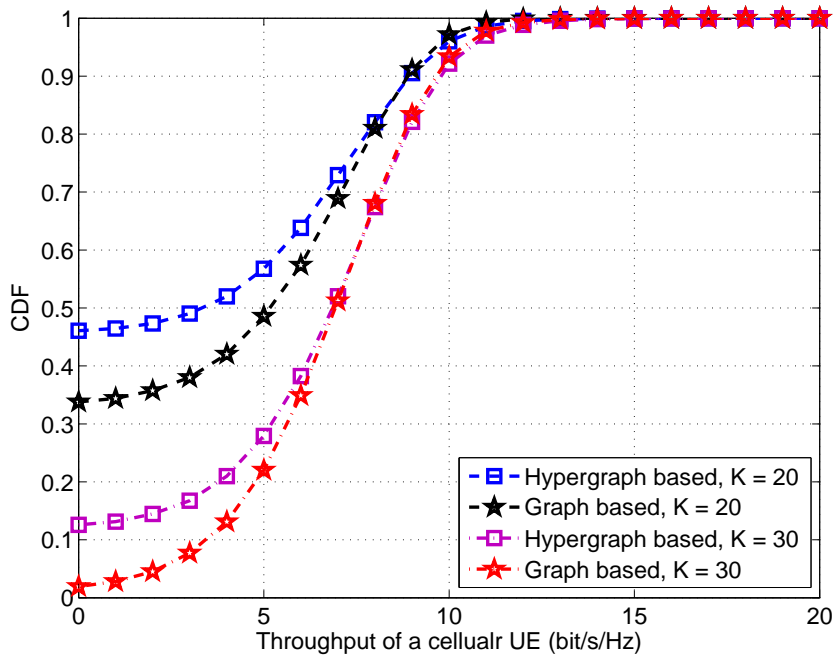}
\caption{Cumulative distribution function of cellular throughput with $N = 30$, and $M = 30$.} \label{CDF_c}
\end{figure}

\begin{figure}[!t]
\centering
\includegraphics[width=5in]{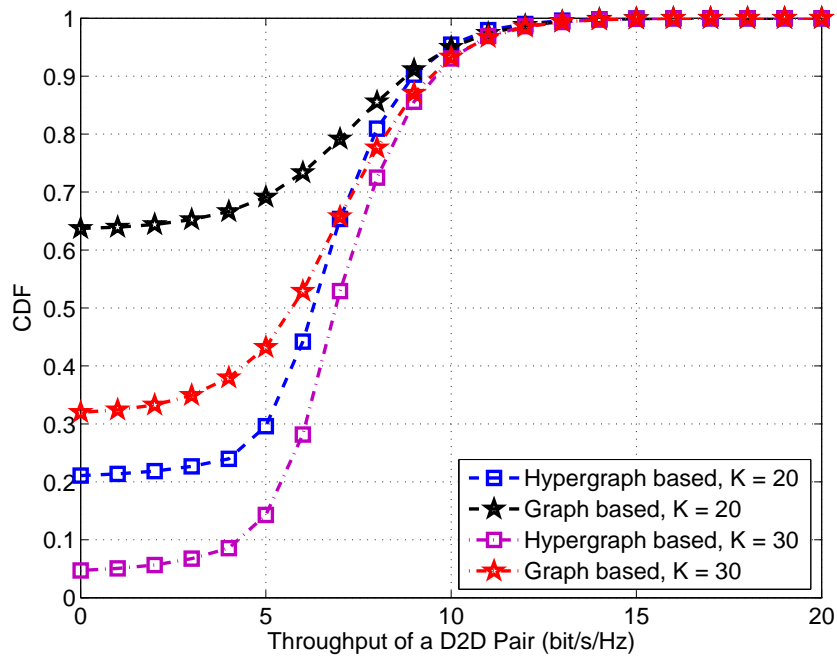}
\caption{Cumulative distribution function of a D2D capacity with $N = 30$, and $M = 30$.} \label{CDF_d}
\end{figure}

\begin{figure}[!t]
\centering
\includegraphics[width=5in]{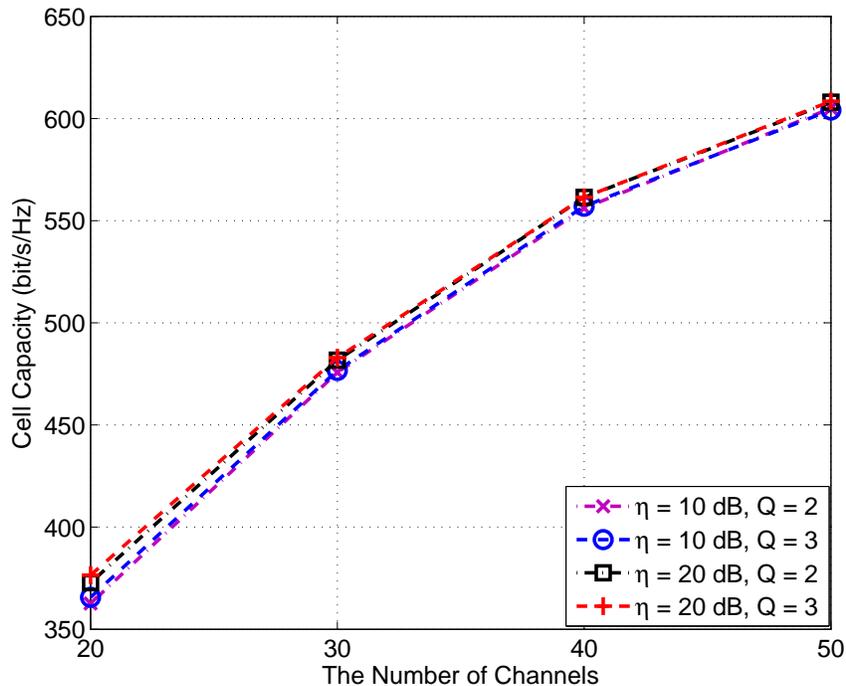}
\caption{Comparison of the cell capacity with the number of channels $K$ for $M = 20$, and $N = 40$.} \label{G}
\end{figure}

\begin{figure}[!t]
\centering
\includegraphics[width=5in]{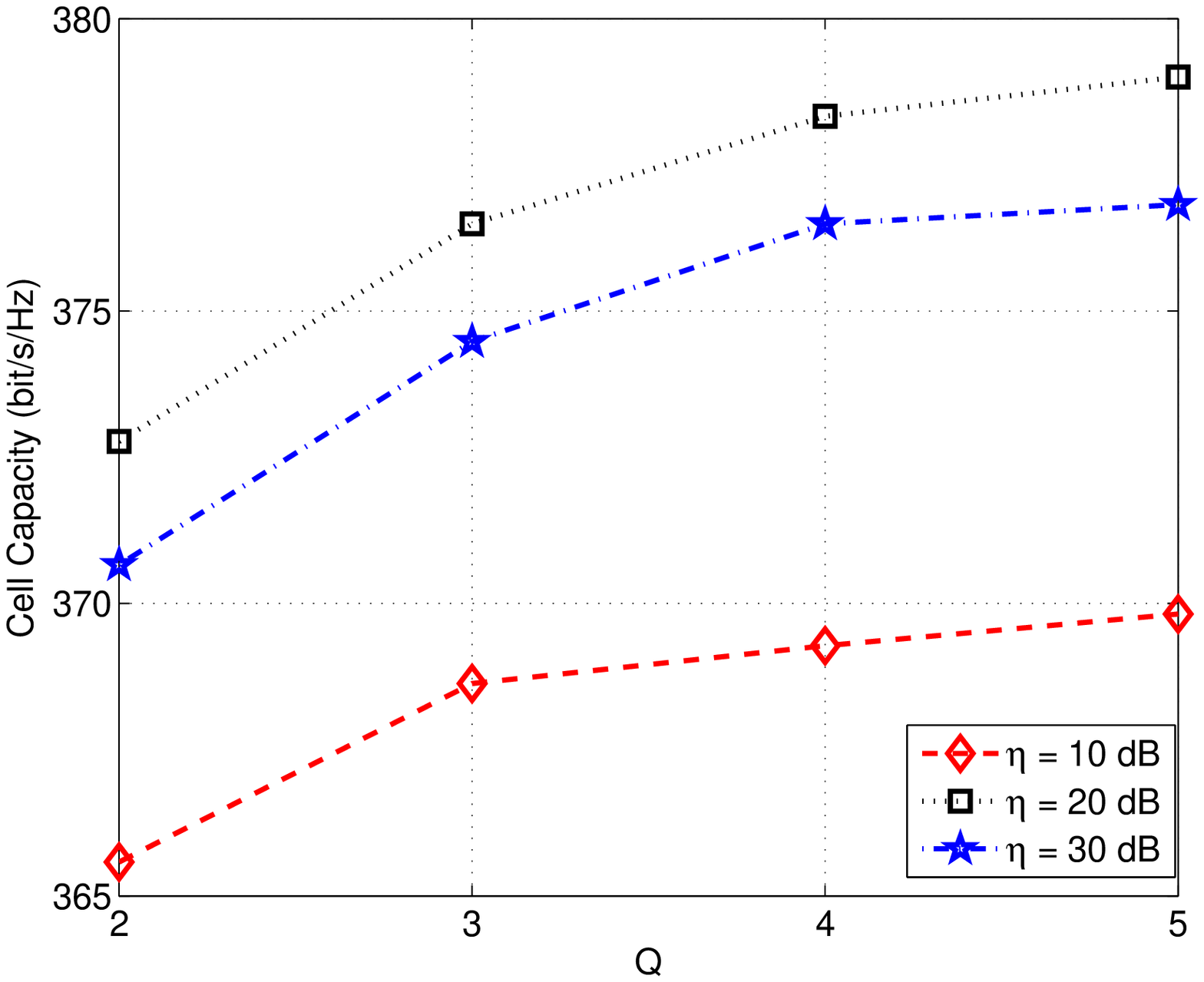}
\caption{Comparison of the cell capacity with $Q$ for $M = 20$, $N = 40$, and $K = 20$.} \label{DG}
\end{figure}

In this section, we present the simulation results of the hypergraph based method in Table \ref{algorithm}, in comparison to the graph based method in Table \ref{graph_algorithm}, and the scenario without D2D, where all the UEs are in the cellular mode. We investigate the relation of the cell capacity to the number of cellular UEs and D2D pairs under two conditions: 1) the number of channels is sufficient for orthogonal access; 2) the number of channels is not sufficient for orthogonal access. For the simulations, we consider a single cell scenario, where cellular communications and D2D communications co-exist, and they can share the channels. The cellular UEs and D2D pairs are distributed randomly in a cell, where the communication distance of each D2D pair cannot exceed a given maximum distance. In this simulation, we use the Shannon capacity model\footnote{In practical system, the signal will be modulated to an OFDM symbol with a certain kind of constellation, such as 64 QAM. Then the receiver will decode this symbol according to the received SINR. The spectrum efficiency is the number of correctly decoded bits per second over the given bandwidth for both the cellular UEs and D2D pairs which use the same channel. In most cases, the received SINR will fall into the linear dynamic range of the decoder. Because of the linear effect, we can obtain a similar result where the value is only rescaled.} to evaluate the cell capacity. In addition, we focus on the frequency domain, and there is no time multiplexing. The simulation parameters are given in Table \ref{parameters}.

In Fig.~\ref{c_cellular}, we show the cell capacity as a function of $N$ cellular UEs with $M = 20$ D2D pairs, and $K = 30$ channels. We can see that the cell capacity with the graph or hypergraph based method increases at first and then decreases. When $N \leq 20$, the cell capacity obtained by the hypergraph based method is almost the same as that obtained by the graph based method, because of low mutual interference. Besides, the cell capacity increases as the number of cellular UEs grows due to the channel sharing. When $N > 20$, the mutual interference becomes large and leads to the decrease in the cell capacity. In addition, for the hypergraph based method, the mutual interference is alleviated by allocating orthogonal channels, since the cumulative interferers are well modeled. Thus, when $N = 50$, the cell capacity obtained by the hypergraph based method is 60 bit/s/Hz higher compared to the graph based method. Compared to the graph based method, the capacity obtained by the hypergraph based method is closer to the optimal result.

Fig.~\ref{c_D2D} illustrates the cell capacity as a function of the number of D2D pairs $M$ with $N = 10$, and $K = 10$. The cell capacity increases as the number of D2D pairs grows, since more D2D pairs are allocated to channels. In addition, it shows that when $M > 40$, the increase of the cell capacity slows down. This indicates that when the number of D2D pairs becomes larger than 40, the cell capacity will be limited by the number of channels. Under the assumption that the UEs in the same edge or hyperedge cannot utilize the same channel, the cell capacity finally becomes saturated, because the number of channels is not sufficient. Simulation results are consistent with \textbf{Proposition 1} in Section \ref{analysis}. When $M = 20$, the cell capacity with the hypergraph based method is about 63 bit/s/Hz higher than that with the graph based method, and the gap becomes 130 bit/s/Hz when $M = 50$. The reason is that when the number of D2D pairs grows, more UEs will share the same channel, leading to larger mutual interference. The hypergraph models cumulative interference with sufficient accuracy, the mutual interference gets alleviated well. Therefore, the gap between the cell capacity using the hypergraph based method and the graph based method increases. From Fig.~\ref{c_cellular} and Fig.~\ref{c_D2D}, if the number of channels is fixed, it can be observed that the effect of cumulative interference modeling is more significant when the number of UEs is larger.

In Fig.~\ref{c_channel}, we provide the cell capacity as a function of the number of channels $K$ with $M = 30$ and $N = 30$. When the number of channels grows, the more UEs can be allocated to channels for communication. Therefore, the cell capacity increases as the number of channels grows. The cell capacity obtained by the hypergraph based method is about 90 bit/s/Hz higher than that obtained by the graph based method when $K = 20$. This implies that the hypergraph can model the interference with sufficient accuracy and hence alleviates it. When $K = 50$, the cell capacity with the graph based method is narrowly close to that with the hypergraph based method. The reason is that the number of channels becomes larger, and hence the number of cumulative interferers decreases.

In Fig.~\ref{CDF_c}, we show the Cumulative Distribution Function (CDF) of the throughput of a cellular UE with $M = 30$, and $N = 30$. Note that we do not use any time time multiplexing, we only provide the throughput in one time interval in this paper. We can observe that the outage probability when $K = 20$ is about 0.3 higher than that when $K = 30$. Under the assumption that different cellular UEs cannot be allocated to the same channel, at least 10 UEs are in outage when $K = 20$, and thus leads to the gap between $K = 20$ and $K = 30$. On average, the cellular UE throughput obtained by the graph based method is 0.8 bit/s/Hz higher than that obtained by the hypergraph based method when $K = 20$. The outage probability with the hypergraph based method is 0.15 higher than that with the graph based method when $K = 20$, which implies that more cellular UEs can be allocated to channels with the graph based method.

Fig.~\ref{CDF_d} shows the CDF of the throughput of a D2D pair with $M = 30$, and $N = 30$. The D2D throughput is about 3.0 bit/s/Hz higher with the hypergraph based method than that with the graph based method when $K = 20$, and 2.1 bit/s/Hz higher when $K = 30$. This shows that the hypergraph based method can effectively improve the throughput of a D2D pair. The outage probability obtained by the graph based method is 0.4 higher than that obtained by the hypergraph based method when $K = 20$, which implies that more D2D pairs can be allocated to channels with the hypergraph based method. Fig.~\ref{CDF_c} and Fig.~\ref{CDF_d} show that when $K = 20$, on average, there are 13.8 cellular UEs, 6.3 D2D pairs outage with the hypergraph based method, and 10.2 cellular UEs, 18.3 D2D pairs outage with the graph based method. We can conclude that more UEs can be allocated to channels with the hypergraph based method when the number of channels is fixed, and hence the spectrum efficiency is improved.

In Fig.~\ref{G}, we compare the cell capacity with different numbers of cumulative interferers in a hyperedge $Q$ and selection thresholds $\eta_c$ and $\eta_d$. Here, we assume that $\eta_c = \eta_d = \eta$.  The cell capacity with $Q = 3$ is about 3 bit/s/Hz higher than that with $Q = 1$ when $K = 20$. Therefore, we can conclude that the cell capacity increases less than 1\% when the value of $Q$ increases. However, the increase of $Q$ will bring significant increase on the computational complexity. The increase of the cell capacity may not make up for the increase in complexity. Therefore, we construct the hypergraph with $Q = 2$.

In Fig.~\ref{DG}, we provide the cell capacity as a function of the value of $Q$ under the assumptions that $M = 30$, $N = 30$, and $K = 30$. As the value of $Q$ increases, more cumulative interferers in a hyperedge would make it easier to form a hyperedge. Therefore, the cell capacity increases because the cumulative interference is well eliminated. Although the cell capacity will increase, the increase might not make up the increase in complexity. Therefore, we construct the hypergraph with $Q = 2$. In addition, with the same value of $Q$, if the threshold becomes high, the cell capacity decreases because the hyperedge will be hard to form. If the threshold becomes low, the number of hyperedges will increase. Under the assumption that the UEs in the same hyperedge cannot use the same channel, fewer UEs will be allocated to channels, and hence the cell capacity decreases.

\section{Conclusions}%
\label{conclusion}
In this paper, we investigate channel allocation by a hypergraph method which coordinates the interference among D2D pairs and cellular UEs in order to increase the cell capacity using D2D underlay communications. We formulate the channel allocation problem as a hypergraph coloring problem to maximize the cell capacity. We also present a greedy coloring algorithm with polynomial complexity proportional to $O((M+N)^3)$, where $N$ and $M$ respectively represent the number of cellular users and D2D pairs. The analysis indicates that proper allocation of D2D pairs can actually increase the cell capacity. The throughput of D2D pairs first increases and then saturates with the increasing number of D2D pairs. Simulation results show that the studied hypergraph based channel allocation method increases the cell capacity by 33\% compared to the traditional graph based method with $N = 50$, $M = 20$ and $K = 30$, where $K$ is the number of available channels.

\end{document}